\documentclass[times]{article}
\usepackage{authblk}
\usepackage{times}
\usepackage[]{natbib}

\usepackage{moreverb,url}
\usepackage[colorlinks,bookmarksopen,bookmarksnumbered,citecolor=red,urlcolor=red]{hyperref}

\usepackage{bm}
\usepackage{graphics}

\usepackage{tikz}
\usetikzlibrary{calc,fit,positioning,arrows,shapes,backgrounds,patterns}
\tikzset{>={latex}}
\definecolor{deepskyblue}{rgb}{0, 0.75, 1}
\tikzstyle{node} =[minimum size = 1cm, text width=1cm, font=\normalsize, align=center]
\tikzstyle{state}=[node, minimum size = 0.8cm, draw, rounded corners, fill=deepskyblue, text width=1.7cm]

\newcommand{\z}{\mathbf{z}}
\newcommand{\btheta}{\mbox{\boldmath $\theta$}}

\begin{document}

\title{A comparison of two frameworks for multi-state modelling, applied to outcomes after hospital admissions with COVID-19}


\author[1]{Christopher H. Jackson}
\author[1]{Brian D. M. Tom}
\author[1,2]{Peter D. Kirwan}
\author[2]{Sema Mandal}
\author[1]{Shaun R. Seaman}
\author[1]{Kevin Kunzmann}
\author[1]{Anne M. Presanis}
\author[1,2]{Daniela De Angelis}

\affil[1]{MRC Biostatistics Unit, University of Cambridge}
\affil[2]{Public Health England}

\date{}

\maketitle

\begin{abstract}
We compare two multi-state modelling frameworks that can be used to represent dates of events following hospital admission for people infected during an epidemic.  The methods are applied to data from people admitted to hospital with COVID-19, to estimate the probability of admission to ICU, the probability of death in hospital for patients before and after ICU admission, the lengths of stay in hospital, and how all these vary with age and gender.    One modelling framework is based on defining transition-specific hazard functions for competing risks.   A less commonly used framework defines partially-latent subpopulations who will experience each subsequent event, and uses a mixture model to estimate the probability that an individual will experience each event, and the distribution of the time to the event given that it occurs. 
We compare the advantages and disadvantages of these two frameworks, in the context of the COVID-19 example.  The issues include the interpretation of the model parameters, the computational efficiency of estimating the quantities of interest, implementation in software and assessing goodness of fit.   In the example, we find that some groups appear to be at very low risk of some events, in particular ICU admission, and these are best represented by using ``cure-rate'' models to define transition-specific hazards. 
We provide general-purpose software to implement all the models we describe in the \texttt{flexsurv} R package, which allows arbitrarily-flexible distributions to be used to represent the cause-specific hazards or times to events.   
\end{abstract}



\section{Introduction}

For emerging infectious diseases such as coronavirus disease 2019 (COVID-19), the severe acute respiratory illness following infection with the SARS-CoV-2 virus, a prompt understanding of disease severity and healthcare resource use are essential to informing government response.  Of particular interest are the probability that a person just admitted to hospital will be admitted to an intensive care unit (ICU), the probability of death in hospital before or after ICU admission, and the predicted length of stay in hospital, or average times between these events.  Accurate estimates of these quantities are required both for direct policy-making, and to include in larger models that combine these estimates with other sources of data (e.g. on transmission and risk of hospitalisation) to predict outcomes for wider populations~\cite{deangelis:hida,birrell:bmcph}.  

These quantities can typically be estimated from data on people hospitalised with the infection.  These data usually consist of dates of admission, discharge, and in-hospital events such as admission to ICU.   The follow-up is commonly incomplete, i.e. it ends while patients are still in hospital, and even during follow-up, the occurrence of events and their dates may be incompletely recorded.   This form of data is generally described as \emph{multi-state} data, in which a set of individuals, sharing the same starting state, are followed up through time.  For each individual, we might observe the time and state of their next transition, or observe that they have remained in the same state up to a particular time. In the latter case, the time of the next transition is right-censored and the state they next move to is unknown.   After a transition is observed, there may be onward transitions to further states observed in the same manner.   

Two different modelling frameworks have been used to represent this kind of data.   The most commonly used method to represent transitions in multi-state data is based on \emph{cause-specific hazards} of competing risks \cite{prentice1978analysis,andersen2002competing,andersen:keiding}. These models are defined by hazards or intensities $\lambda_{rs}(t)$ representing the risk of transition to state $s$ for a person in state $r$ at a time $t$.  These can be interpreted as defining distributions for latent times $T_s$ to competing events $s$ for an individual, with the minimum of the $T_s$ defining the transition that happens.   Putter et al. \cite{putter:mstate} present a tutorial for nonparametric and semi-parametric competing risks and multi-state models, while Crowther and Lambert \cite{crowther2017parametric} describe a flexible framework for multi-state modelling based on a range of parametric distributions.  Ieva et al. \cite{ieva2017multi} present a case study of multi-state modelling of hospital admission data using both semi-parametric and Weibull parametric models.

A less common approach is based on mixture models. This was used by Ghani et al. \cite{ghani2005methods} and Donnelly et al. \cite{donnelly2003epidemiological} in a context that is similar to our application, to estimate probabilities of death and recovery, and times to these events, for people with an infectious disease.  Each individual's next state is assumed to be $s$ with probability $p_s$, where $s=1,2$ represent death and recovery, so that $p_1$ is the case fatality ratio, and $p_1+p_2=1$.  Then for each $s$, a parametric model is specified for the time $T_s$ until the event, given that $s$ is the event that occurs.  Since the next state is unknown for individuals whose transition time is right-censored, the likelihood for the $p_s$ and the parameters of the time-to-event models takes the form of a mixture model with partially-known component membership.   This model was described in more generality by Larson and Dinse \cite{larson1985mixture}, who represented $S$ states, and included covariates both on the event probabilities and on the models for times to events.  These models include our quantities of interest as explicit parameters, and have been used with flexible component-specific distributions \citep{lau2011parametric}, though have not been used, as far as we know, to represent multi-state models with onward transitions following the first transition. 


These approaches to competing risks were contrasted by Cox \cite{cox1959analysis} (with two events and exponentially-distributed times) who suggested either approach could be used to construct an arbitrarily-flexible model.  However they have not been compared in practice in the context of general fully-parametric multi-state models.    Lau et al. \cite{lau2009competing} advocated mixture models, compared to semiparametric proportional cause-specific hazard models, for describing the associations of covariates with the risk of competing outcomes, but did not consider fully-parametric cause-specific hazard models.   In our application, we focus on parametric models, to stabilise estimation in periods where the data are sparse, enable short-term extrapolation beyond the end of the data, and to provide explicitly parametric inputs for epidemic models, based on Bayesian evidence synthesis, that are designed to inform policy for wider populations~\cite{deangelis:hida,birrell:bmcph}.


In this paper, we compare the practical use of these two different frameworks for fully-parametric multi-state modelling, in the context of an application to COVID-19 hospital admissions.   We consider their interpretation, the ease of model specification, their computational efficiency, implementation in software, and assessing their goodness of fit to the data.  A novel modification to the likelihood in both frameworks was required to represent partially-observed final outcomes, where patients are known to be alive but with unknown hospitalised status.    We develop the first general-purpose software implementation of the mixture multi-state model, in the \texttt{flexsurv} package for R \cite{flexsurv}, and facilities in that package
for multi-state models with cause-specific hazards are extended, e.g. to handle different distribution families for each transition \cite{crowther2017parametric}.

A feature of our data is that there are subsets of patients who appear to be at very small risks of particular  events.   These are handled naturally by the mixture model, which is parameterised by probabilities of events.  To characterise these data within the cause-specific hazards framework, we use \emph{cure} (also known as \emph{mixture cure}) distributions to represent one or more of the latent event times $T_k$.   Originating from Boag \cite{boag1949maximum}, these are used to describe populations with a disease (typically cancer) where a latent proportion are cured and never die from the disease.   They are typically fitted to data where either the time of death or a right-censoring date is observed, but cure or time to cure cannot be observed directly \cite{farewell1982use,jakobsen2020estimating}.   The cured fraction and distribution of the time to death among the non-cured fraction are estimated jointly.   As in Conlon et al.  \cite{conlon2014multi}, we use cure distributions as cause-specific hazards in a competing risks model.  Note the distinction from the mixture multi-state model of Larson and Dinse \cite{larson1985mixture}, in which the \emph{time} of achieving ``immunity'' from death  (e.g. recovery or cure) \emph{can be observed}, and the model aims to describe this time, as well as the probability of the event.  While Ghani et al. \cite{ghani2005methods} referred to their model as a ``cure'' model, it is an example of the mixture multi-state model rather than the Boag \cite{boag1949maximum} model, since both competing events (death and recovery) were observable. 


In Section~\ref{sec:chess} we describe the motivating COVID-19 hospital admissions dataset.   Section~\ref{sec:models} sets out the theoretical definitions of the two alternative multi-state modelling frameworks and how the quantities of interest are defined and computed.   The models are applied to the COVID-19 hospital data in Section~\ref{sec:results}, and the fit and interpretations of the best of both types of parametric model are compared.    While the mixture model is simpler to interpret and involved less computation to obtain the the quantities of interest, the fit in the COVID-19 example was best for the cause-specific hazards model with mixture-cure distributions.   Section~\ref{sec:discussion} concludes the paper with a discussion of the merits of the two frameworks and the open issues.

\section{CHESS: hospital admissions data from COVID-19 patients \label{sec:chess}}

Data were extracted from the COVID-19 Hospitalisation in England Surveillance System (CHESS), established and managed by Public Health England (PHE) \citep{phe:chess}. CHESS began in mid-March 2020 and aims to monitor the impact of severe COVID-19 infection on the population and on health services and provide real-time data to forecast and estimate disease burden and health service use. CHESS is a mandatory data collection system \citep{phe:sgss} and captures individual-level data on all patients admitted to ICU or HDU (high dependency unit) with COVID-19 at National Health Service (NHS) Trusts in England, in addition to data from all hospital admissions with COVID-19 from the 22 out of 107 trusts that were designated as ``sentinel'' trusts, representing 26\% of hospital admissions. Collected data include patient demographics, risk factors, clinical information on severity, and outcome. CHESS data covering the period from 15 March to 2 August 2020 were extracted at PHE and linked using NHS number to the Office for National Statistics (ONS) deaths register to obtain complete information for deaths occurring both within and outside of hospital.

There are 5544 hospital admission records from sentinel trusts in CHESS from people known to have tested positive for COVID-19 infection up to 2 days after admission, who are assumed to have been infected outside hospital.   People who were infected with COVID-19 while in hospital for another condition are excluded, as there is insufficient information to determine how much of their hospital stay was due to COVID-19. From the 5544 records, we excluded 74 patients, including those with missing or inconsistent dates of ICU admission, or missing information on age or gender, leaving 5470 hospital admissions.    The final outcomes from these patients, and whether they were admitted to ICU, are counted in Table~\ref{tab:counts}.    In around 10\% of cases, whether the patient died in hospital care, or was discharged, is not yet observed or unknown, noting that people recorded as ``transferred'' are still in hospital care.    The people with ``unknown'' final outcome are known to be still alive, and it is assumed that if they went to ICU then this was recorded, but it is not known whether they are still in hospital at the date of data extraction. 

\begin{figure}[htbp]
\includegraphics[width=\textwidth]{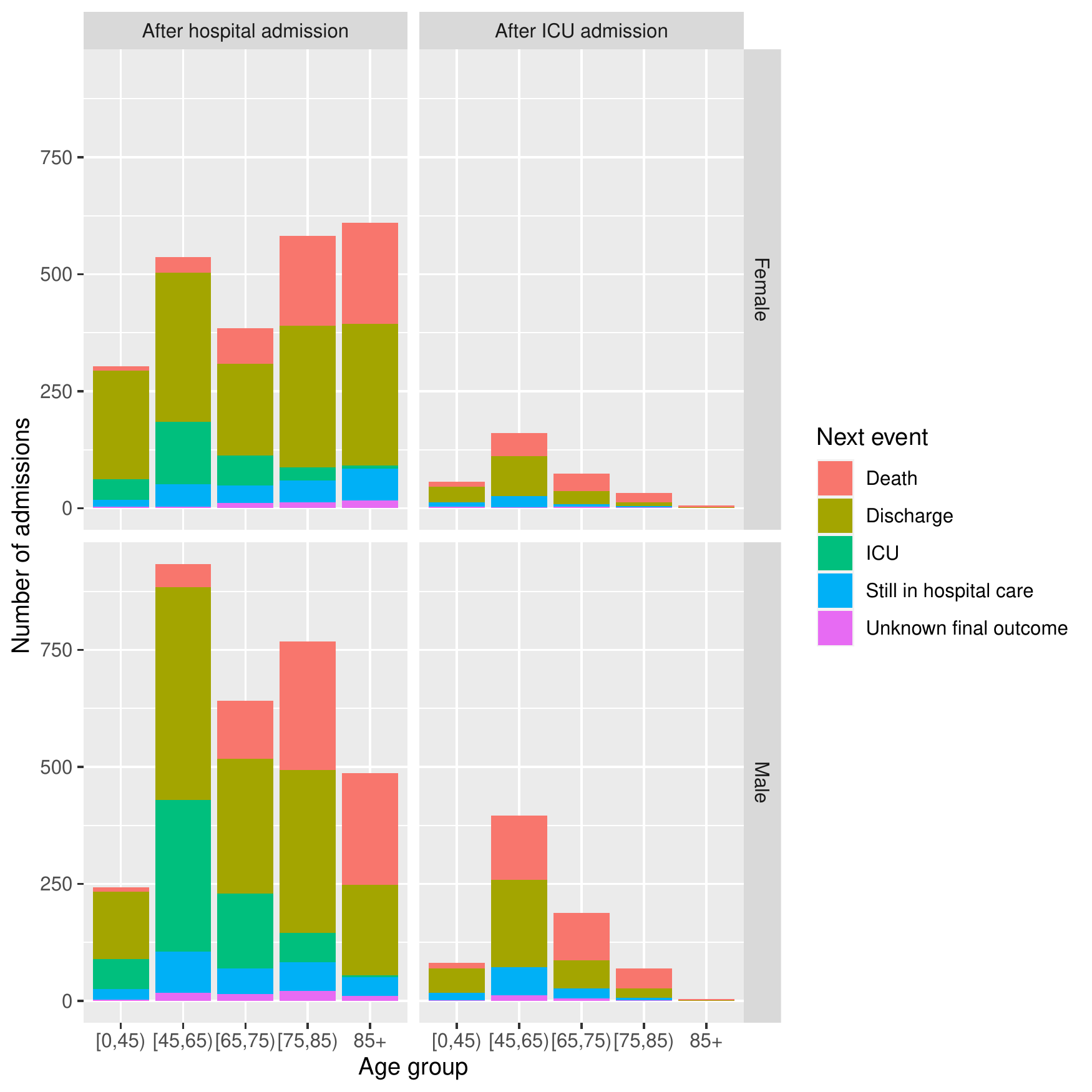}  
\caption[Next state probabilities]{Distribution of next event following hospital admission, or following ICU admission, by age group and gender, as a simple summary of the data.}
\label{fig:barplot}
\end{figure}

\begin{figure}[htbp]
  \includegraphics[height=0.48\textheight]{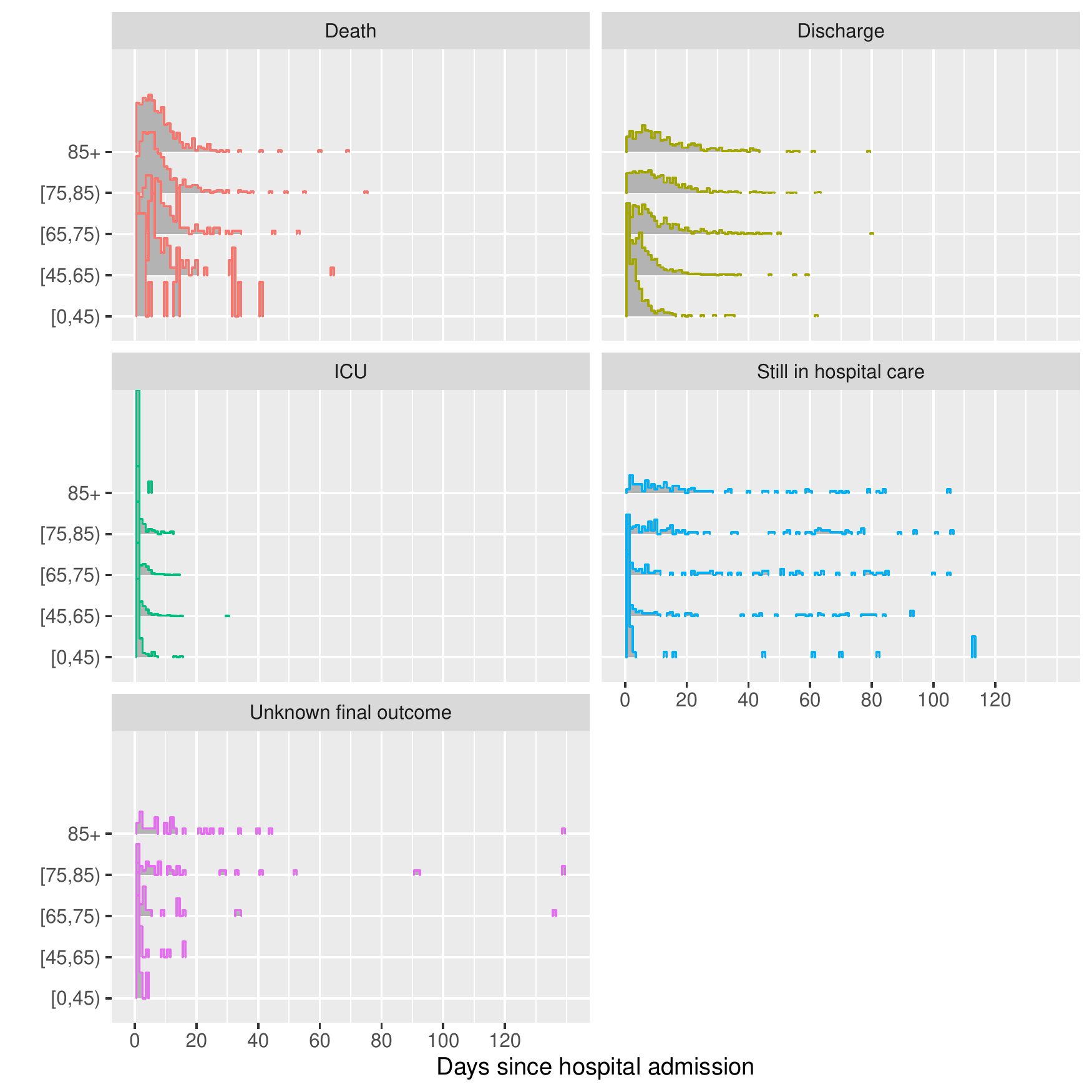}
  \includegraphics[height=0.48\textheight]{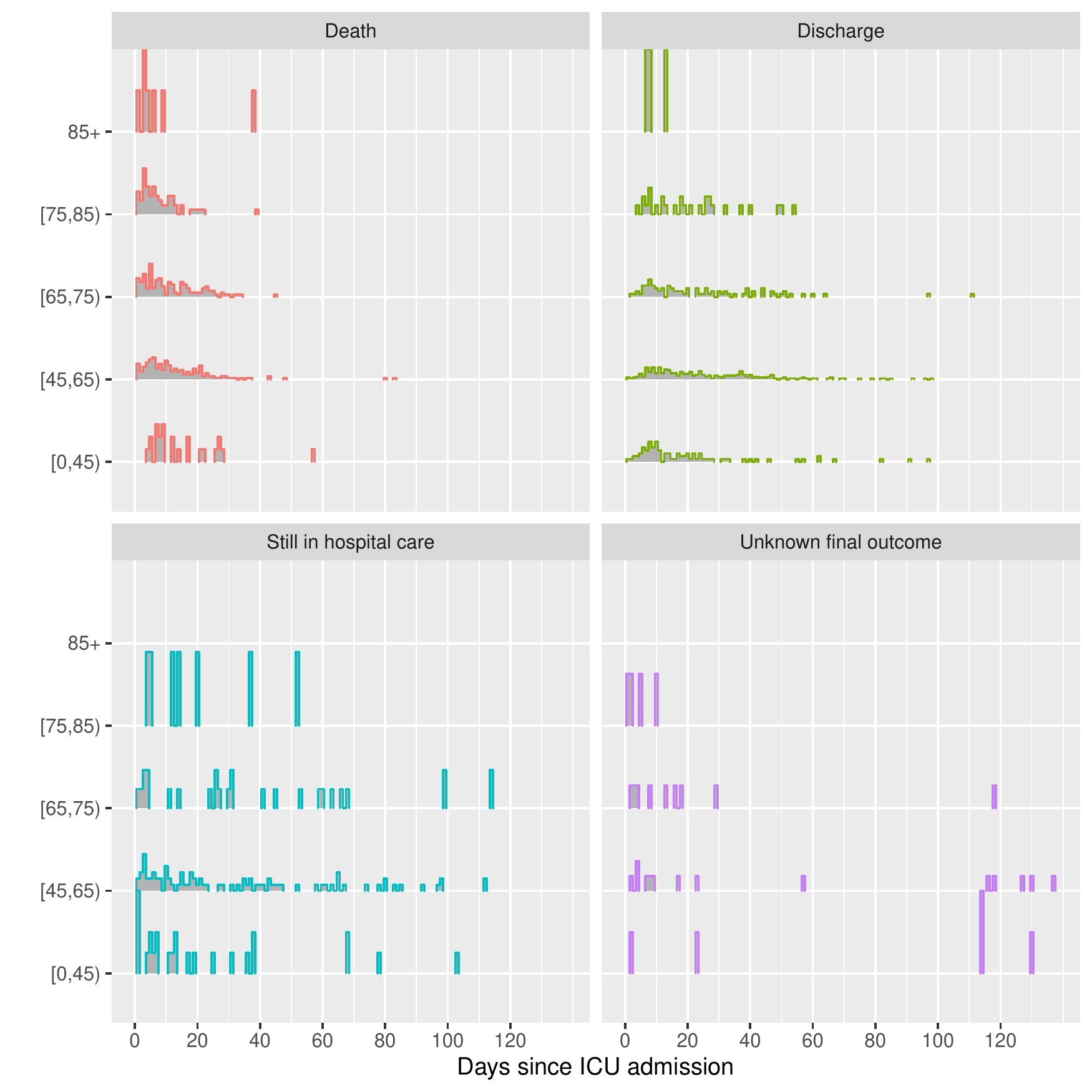}
  \caption[Times to events]{Distribution of times to next event following hospital admission, by age group and gender, as a standard histogram of the data, including times to right-censoring for those still in hospital, or times to data extraction for those with unknown outcome.}
\label{fig:ridge}
\end{figure}

The data on the next event following hospital admission are illustrated further by age and gender in Figure~\ref{fig:barplot} and Figure~\ref{fig:ridge}.  Figure~\ref{fig:barplot} illustrates the proportions experiencing each next event, while Figure~\ref{fig:ridge} shows the distribution of the times to each kind of observed event after admission, the times to right-censoring for those known to be still in hospital, and the times from admission to data extraction for those with unknown outcome.   Note that a substantial number of people are still in hospital care after around 50 days in hospital, which is extreme compared to the distributions of the times to the observed events.   The probabilities and average times of events could be inferred directly from these summaries, but the estimates would be biased as they would ignore the information from the censored times, which indicate longer hospital stays, and people with unknown outcome.   Therefore likelihood-based statistical models are constructed.

\begin{table}

\begin{tabular}[htbp]{lp{0.3in}p{0.6in}p{0.7in}p{0.6in}p{0.6in}p{0.5in}}
\hline
Admitted to ICU  &Died&Discharged&Still in unit&Transferred &Unknown final outcome&Total\\
\hline
Yes & 	418 & 	476 & 	39 &	104 &	32 & 	1069  (19\%)\\
No	& 1221 & 	2780 & 	138 &	198 & 83 & 	4420  (81\%)\\
Total	& 1639 (30\%) &	3256 (59\%)	&177 (3\%)  &	302 (6\%) & 115 (2\%) &5489\\
\hline
\end{tabular}
\caption{Summary of events in CHESS data \label{tab:counts}}
\end{table}

\section{Multi-state modelling frameworks}
\label{sec:models}

A parametric, continuous-time, semi-Markov multi-state model is used with four states, represented in Figure~\ref{fig:multistate}.   This comprises two submodels  representing ``competing risks'' of the next event: one for the event following hospital admission, and another for the event following ICU admission.  Note that in practice, individuals will be discharged from ICU to a hospital ward,  but since dates of leaving ICU are only partially recorded in this dataset, we simply combine the ICU state with the state of being in a hospital ward after leaving ICU, so the ``ICU'' state represents ``in hospital, and has been admitted to ICU'', rather than ``in hospital and currently in ICU''.   Individual $i$'s transition intensity from state $r$ to state $s$, at a time $t$ after entering state $r$, is denoted by $\lambda_{i,r,s}(t)$.   Note the common Markov assumption has been relaxed, thus the intensities are allowed to depend on how long a person has spent in the current state.   Two alternative model parameterisations are used, termed \emph{cause-specific hazards} and \emph{mixture model} formulations.   These differ in how the transition intensities are defined, but they can both be used to estimate our principal quantities of interest, which are:

\begin{enumerate}

\item The \emph{next-state probability} $\pi_{rs}$: the probability that the next state visited after state $r$ is state $s$. 

\item The distribution of the \emph{conditional length of stay} $S_{r,s}$, the time to state $s$ for a person who has just entered state $r$, given that a transition to state $s$ occurs.  The mean of this distribution might be used to describe expected total hospital resource usage for a population.  The variability between individuals in length of stay could be described by quantiles of the distribution of $S_{r,s}$. 

\item The \emph {ultimate-outcome probabilities} $\pi^{(U)}_s$, the probability that the the ultimate outcome for a person just admitted to hospital is state $s$, which can only be death or discharge (the \emph{absorbing states} of the model).

\item The distribution of the \emph{conditional time to ultimate outcome} $U_{s}$, the time to an ultimate outcome $s$ of death or discharge, given that this outcome occurs. 

\end{enumerate}

In general, any summary of the distribution of state transitions and times to events can be computed by simulation, but some quantities will be available analytically. 

\tikzstyle{fromhosp}=[color=gray,very thick,dashed]
\tikzstyle{fromicu}=[color=black,very thick,dotted]

\begin{figure}[htbp]
\begin{tikzpicture}
\node (hospital) [state] {1. Hospital};
\node (icu) [state, right = of hospital] {2. ICU};
\node (finalanchor) [right = of icu ] {};
\node (discharge) [state, above = of finalanchor ] { 3. Death };
\node (death) [state, below = of finalanchor ] { 4. Discharge };
\draw [->,fromhosp] (hospital) -- (icu); 
\draw[->,fromhosp] (hospital) -- (discharge);
\draw[->,fromicu] (icu) -- (discharge);
\draw[->,fromhosp] (hospital) -- (death);
\draw[->,fromicu] (icu) -- (death);
\draw[->,fromhosp] (6.5,-0.5) -- (7.5,-0.5); \\
\draw[->,fromicu,] (6.5,   -1) -- (7.5,-1); \\
\node[align=left] at (8, 0) {Submodels};
\node[align=left] at (9, -0.5) {From hospital};
\node[align=left] at (9, -1) {From ICU};
\end{tikzpicture}

\caption{Multi-state model. Permitted instantaneous transitions between states indicated by arrows.}
\label{fig:multistate}
\end{figure}
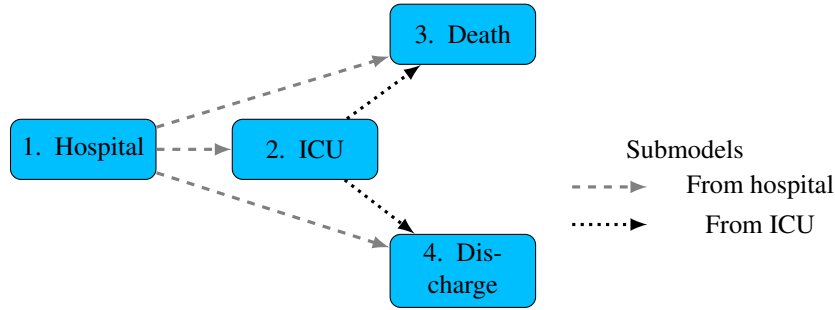

\subsection{Cause-specific hazards model}

In a competing risks model specified through cause-specific hazards\cite{prentice1978analysis,andersen:keiding,putter:mstate} at time $t$ after entering state $r$, an individual is subject to a risk of transition to any state $s$ allowed by the model structure, so that $\lambda_{i,r,s}(t) > 0$ only if $s$ is in the set $\mathcal{S}_r$ of potential next states after state $r$  (indicated by the arrows in Figure \ref{fig:multistate}).   For example, a person currently in ICU could either die or be discharged at any time, but may not return to the ``hospital'' state.   The transition intensity $\lambda_{i,r,s}(t) $ can be interpreted as the hazard function of a parametric distribution that governs a (latent) time $T_{r,s}$ from state $r$ entry until the transition to state $s$.  Only one of these competing transitions will actually happen, that is, the one that happens first, to state $s^{(*)}_r  = arg min_{u \in \mathcal{S}_r} \{T_{ru}\}$.  In this formulation, the quantities of interest are defined as follows.

\paragraph{Conditional length of stay} This is
\[ S_{rs} = (T_{rs} | s^{(*)}_r = s) = (T_{rs} | min_{u \in \mathcal{S}_r} \{T_{ru}\} = T_{rs})\]
the random variable whose distribution is the conditional distribution of $T_{rs}$ given that the transition from state $r$ to state $s$ occurs.    

\paragraph{Next-state probabilities} $\pi_{rs}$ are determined by considering the implicit ``competing risks'' submodels that define the full multi-state model, one submodel governing each state that a person can
transition from.  In Figure~\ref{fig:multistate} there are two submodels, one governing the next state after hospital admission, and one for the next state after ICU admission.  These are indicated by dashed lines and dotted lines respectively.   In the hospital submodel, death before ICU, ICU admission and discharge before ICU are treated as ``absorbing'' states (which individuals cannot leave), while in the ICU submodel, the absorbing states are death and hospital discharge folllowing ICU admission.   Then the \emph{continuous-time transition probabilities} of the full multi-state model and a particular submodel (labelled $M$) are defined respectively as:
\begin{itemize}

\item $p_{r,s}(t)$, the probability that an individual in state $r$ occupies state $s$ at a time $t$ after entering state $r$

\item $p^{(M)}_{r,s}(t)$, the probability that an individual currently in state $r$ of submodel $M$ is in state $s$ of the submodel at at time $t$ later --- i.e. the probability that a person has experienced the competing risk $s$, or, that they have entered the equivalent state of the full multi-state model some time in the past.
\end{itemize}
In the COVID-19 example, we have probabilities $p^{(H)}_{r,s}(t)$, $p^{(I)}_{r,s}(t)$ for the states at time $t$ after hospital and ICU admission within the hospital and ICU submodels respectively.  The next-state probabilities after $r =$ hospital are then $\pi_{r,s}(t)  = lim_{t \rightarrow \infty} p^{(H)}_{r,s}(t)$, and the probabilities for the next state after $r=$ ICU are $\pi_{r,s}(t)  = lim_{t \rightarrow \infty} p^{(I)}_{r,s}(t)$. 

\paragraph{Ultimate outcomes}
Similarly, the ultimate-state probabilities are determined from the transition probabilities of the full multi-state model as $\pi^{(U)}_s = lim_{t \rightarrow \infty} p_{r,s}(t)$ for $r=$ Hospital. 
The conditional time $U_{s}$ to an ultimate outcome $s$, conditionally on this outcome occuring, is defined by first identifying all pathways $\mathcal{P}$ through the states that end in outcome $s$.  In this example, there are only two pathways from hospital to death or discharge: where ICU was visited or wasn't visited.  $U_s$ is then defined by calculating the total time taken to travel each pathway, $U_s^{(\mathcal{P})} = \sum_{(r,s^\prime) \in \mathcal{P}} S_{r,s^\prime}$, then averaging with respect to the probability of each pathway: $U_s = \sum_{\mathcal{P}} U_s^{(P)} Pr(\mathcal{P})$.

\paragraph{Computation of quantities of interest}

The next-state probabilities are calculated as follows.   The transition probabilities for the two sub-models are related to the transition intensities through the Kolmogorov forward equation: 
\[ \frac{dP^{(H)}(t)}{dt} = P^{(H)}(t)Q^{(H)}(t), \quad \frac{dP^{(I)}(t)}{dt} = P^{(I)}(t)Q^{(I)}(t)\]
where $P^{(H)}(t)$, $P^{(I)}(t)$ are the matrices with $r,s$ entry $p^{(H)}_{r,s}(t)$, $p^{(I)}_{r,s}(t)$, respectively,
and  $Q^{(H)}(t)$ and $Q^{(I)}(t)$ are the \emph{transition intensity matrices} of the hospital and ICU submodels, with $r,s$ entry given by $\lambda_{i,r,s}(t)$ for $r \neq s$, where $r$ and $s$ index the states in each submodel.  The diagonal entries of the intensity matrix are then defined so that the rows sum to zero.  The initial condition for the equation in each case is a transition probability matrix defined by the identity matrix at $t=0$.  The Kolmogorov forward equation is solved numerically using the R package \texttt{deSolve} \citep{deSolve}, giving estimates of all $p^{(H)}_{r,s}(t)$ and $p^{(I)}_{r,s}(t)$, hence (by choosing a very large value of $t$) estimates of $\pi_{r,s}$. 

The distribution of the conditional lengths of stay $S_{r,s}$, the ultimate-outcome probabilities $\pi^{(U)}_s$, and the distribution of the conditional times to ultimate outcome $U_{s}$, are all determined by simulation.   State transition histories (until death or discharge) are simulated from the fitted model for a large population, and the resulting sample is summarised to give the quantities of interest.   The next event and the time to this event for a person in state $r$ is simulated by generating latent times $T_{rs}$ from each cause-specific distribution $s$ implied by the hazards $\lambda_{rs}(t)$, with the minimum of the competing times defining the one that happens.

\subsection{Mixture multi-state model}

In the ``mixture'' competing risks model \citep{larson1985mixture}, extended here to a general multi-state model, each individual $i$ in state $r$ makes a transition to a destination state $s$ that is determined randomly at time 0.  Thus their transition intensity at time $t$ after state $r$ entry is defined by
  \[
    \lambda_{i,r,s}(t)  = \left\{
                \begin{array}{ll}
                  \lambda^*_{i,r,s}(t) & \mbox{if } I_{i,r} = s \\
                  0                           & \mbox{if }  I_{i,r} \neq s\\ 
                \end{array}
              \right.
  \]
where $I_{i,r}$ is a latent categorical variable that determines which transition will happen next for an individual $i$ in state $r$, governed by probabilities $\pi_{r,s} = P(I_{i,r} = s)$, with $\sum_{s \in \mathcal{S}_r} \pi_{r,s} = 1$.    The transition intensity $\lambda^*_{i,r,s}(t) $ is defined by the hazard function of a parametric distribution that governs the time $S_{r,s}$ from state $r$ entry until the transition to state $s$, \emph{given that this is the transition that occurs}.  Unlike in the cause-specific hazards model, we do not model the times to events that don't occur.

\subsection{Data and likelihoods}

Observations of the event times or censoring times from individual $i$ are indexed by $j$.   Each observation is one of three types, indicated by a ``status'' $\delta_{i,j}$. 

\begin{itemize}
\item \emph{exact transition time:}  $\mathbf{Y}_{i,j} = \{y_{i,j}, r_{i,j}, s_{i,j}, \delta_{i,j}=1\}$, where a transition to state $s_{i,j}$ is known to occur at a time $y_{i,j}$ after entry to state $r_{i,j}$. 

\item \emph{right censoring}  $\mathbf{Y}_{i,j} = \{y_{i,j},r_{i,j},\delta_{i,j}=2\}$, where an individual's follow-up ends while they are in state $r_{i,j}$, at time $y_{i,j}$ after entering this state, thus the next state and the time of transition to it are unknown. 

\item \emph{partially-known outcomes} $\mathbf{Y}_{i,j} = \{y_{i,j},r_{i,j},\delta_{i,j}=3\}$, from individuals known to be alive, where it is assumed known whether or not they went to ICU, but it is unknown whether they are still in hospital at time $y_{i,j}$ after entry to state $r_{i,j}$, or were discharged at some time before $y_{i,j}$.

\end{itemize}

\paragraph{Cause-specific hazards model} 

We use the likelihood as in Prentice et al. \cite{prentice1978analysis}, extended to handle partially-known outcomes.  To construct this, first write the $r \rightarrow s$ transition intensity for individual $i$ as $\lambda_{i,r,s}(t) = h_{r,s}(t | \btheta, \z_i) $, where $\z_i$ is a vector of individual-specific, time-constant covariates, $\btheta$ is a vector including all parameters of the parametric distribution with hazard of the form $h_{r,s}(t)$ and effects of covariates on the parameters,  and $\btheta_{r,s}$  indicates the specific elements of $\btheta$ that pertain to the $r,s$ transition.  The likelihood is constructed without reference to latent event times $T_{rs}$, thus does not assume, e.g. that these are independent.  Denote the corresponding probability density function and cumulative distribution function as $f_{r,s}(t), F_{r,s}(t)$ respectively (omitting the conditioning for clarity).   Then the $j$th observation from individual $i$ contributes the following term $l_{i,j}$ to the likelihood: 

\begin{itemize}

\item For exact transition times, $\delta_{i,j}=1$,

  \[ l_{i,j} = f_{r_{i,j},s_{i,j}}(y_{i,j} | \btheta_{r_{i,j},s_{i,j}}) \prod_{u \in \mathcal{S}_{r_{i,j}}, u \neq s_{i,j}} \big(1 - F_{r_{i,j},u}(y_{i,j} | \btheta_{r_{i,j},u})\big).\]

The first term represents the transition that was observed at $y_{i,j}$, and the second term represents the knowledge that the individual was at risk of transition to the competing states $u$, but these transitions didn't happen by time $y_{i,j}$.  In other words, there are right-censored times of transition to each of these states $u$.

\item 
  For observations of right-censoring, $\delta_{i,j}=2$,

  \[  l_{i,j} =\prod_{u \in \mathcal{S}_{r_{i,j}}} \big(1 - F_{r_{i,j},u}(y_{i,j} | \btheta_{r_{i,j},u})\big) \]

 representing the knowledge that the individual was at risk of all potential transitions from state $r_{i,j}$, but none of these happened by time $y_{i,j}$. 

 \item For individuals with partially-known outcomes, where it is known they are alive and it is known whether they went to ICU or not, the times to death or discharge are right-censored.  Since we do not know whether they have been discharged before time $y_{i,j}$, or if they are still in hospital (thus with discharge time right-censored at $y_{i,j}$), they do not provide any information about the distribution of (potential) discharge times.  Therefore, their likelihood contribution is

  \[  l_{i,j} =1 - F_{r,u}(y_{i,j} | \btheta_{r,u}) \]

 where $r$ denotes either a transition from hospital or a transition from ICU, and $u=$ death.

\end{itemize}

The full likelihood is $l(\btheta | \mathbf{Y}) = \prod_{i,j}l_{i.j}(\btheta | \mathbf{Y}_{i,j})$ over all individuals $i$ and observations $j$, where  $\mathbf{Y}$ is the complete data.   To facilitate computation, however, we write the full likelihood as a product of terms specific to each $r,s$ transition \cite{andersen:keiding}.

\[ l(\btheta | \mathbf{Y}) = \prod_{r,s}l^*_{r,s}(\btheta_{r,s} | \mathbf{Y}) \]
This is possible since, for each $r,s$ transition, we use parametric models with distinct parameters $\btheta_{r,s}$ (and potentially also different distributional forms).   The full multi-state model can then be fitted by maximising each of the transition-specific likelihoods independently, using separate calls to a survival modelling function.

\paragraph{Mixture multi-state model} 

The likelihood presented by Larson and Dinse~\cite{larson1985mixture} for competing risks can be extended easily to a full multi-state model.  Firstly define again $\pi_{r,s}$ as the probability that the next transition for someone in state $r$ is to state $s$.  Then we specify a parametric distribution with density $f_{r,s}(|\theta_{r,s})$  (and  CDF $F_{r,s}()$)  for the time of transition to state $s$ for a person in state $r$, conditionally on this transition being the one that occurs.   Therefore, for an exact time of transition to a known state $s_{i,j}$, i.e.  $\delta_{i,j}=1$, the likelihood contribution is simply 

  \[ l_{i,j} = \pi_{r_{i,j}s_{i,j}} f_{r_{i,j},s_{i,j}}(y_{i,j} | \btheta_{r_{i,j},s_{i,j}}) \] 

For observations $j$ of right-censoring at $y_{i,j}$, the state that the person will move to, and the time of that transition, is unknown.  Thus it is unknown which of the distributions $f_{r,s}$ the transition time will obey, and the likelihood contribution is of the form of a mixture model: 

  \[ l_{i,j} = \sum_{s \in \mathcal{S}_r} \pi_{r_{i,j}s} (1 - F_{r_{i,j},s}(y_{i,j} | \btheta_{r_{i,j},s})) \] 

For patients with partial outcomes, the likelihood is as for right-censoring, except that the term for $s=$ Discharge is simply $\pi_{r_{i,j},s}$, as there is no information about discharge times for these patients. 
  
The full likelihood is $L(\bm{\pi}, \btheta | \mathbf{Y}) = \prod_{i,j} l_{i,j}$, where $\bm{\pi}$ includes all the next-state probabilities $\pi_{r,s}$, and $\btheta$ includes the parameters of the time-to-transition distributions, which may include the effects of covariates.   The $\pi_{r,s}$ may further depend on covariates, e.g. through a multinomial logistic regression model \cite{larson1985mixture}.

\subsection{Parametric distributions and implementation}

The cause-specific hazards models can be fitted in standard survival modelling software.  We use the \texttt{flexsurv} package in R, extending it to handle different distribution familes for different transitions. 
Similar facilities are available in Stata \cite{crowther2017parametric,crowther2014general}.   The \texttt{flexsurv} package was also extended here to implement the likelihood for the mixture multi-state model.   Note that the mixture likelihood needs to be maximised over many more parameters than the likelihood of a comparable cause-specific hazards model, since it does not factorise into independent terms for each component.   An EM algorithm \cite{larson1985mixture} was used to maximise the likelihood, which was found to be more efficient than direct maximisation.   An implementation of the mixture model, restricted to two competing events and with a limited choice of distributions, is also available in the R package \texttt{RISCA} \cite{RISCA}. 

Each model formulation requires the choice of a parametric model for the time to an event.  A wide range of flexible distributions are used in practice and are available in \texttt{flexsurv}, which also allows users to implement new distributions.  Covariates may be included on any parameter of any distribution through a linear model (on the log scale if the parameter is defined to be positive).  A particularly useful form is the generalized gamma distribution, which, in the parameterisation from Prentice \cite{prentice:loggamma}, includes the log-normal, Weibull and gamma distributions as special cases.  The cumulative distribution function is
\[ F(t|\mu,\sigma,Q) =  
\begin{array}{ll}
F_G(\exp(Qw)/(Q^2) ~ | ~   1/Q^2, 1)  &   (Q > 0 )\\
1 - F_G(\exp(Qw)/(Q^2)  ~ |  ~  1/Q^2, 1)  &   (Q < 0)\\
F_L(t ~ | ~ \mu, \sigma) &  (Q = 0)\\
\end{array}
\]
where $w = (\log(t) - \mu)/\sigma$, $F_G(t | a,b)$ is the CDF of the gamma distribution with shape $a$ and rate $b$, $F_L(t |\mu,\sigma)$ is the CDF of the log-normal distribution with log-scale mean $\mu$ and standard deviation $\sigma$, $\mu,Q$ are  unrestricted, and $\sigma$ is positive. 

For the data in this example, we also use \emph{mixture cure} distributions \cite{boag1949maximum}.  These are defined by extending a standard parametric time-to-event distribution $F(t|\theta)$ to include a probability $p$ that the event never occurs, obtaining a model 

\[ Pr(T \leq t | p,\theta) = (1-p)F(t|\theta) \]
This model may be extended to include covariates that explain the ``cure'' probability $p$, through logistic regression.    The \texttt{flexsurvcure} package \cite{flexsurvcure} is used to facilitate the implementation of the cure model in \texttt{flexsurv}. 



\subsection{Model selection}
 
For both the mixture model and the cause-specific hazards / competing risks model, a well-fitting parametric model was selected by the following procedure, on the basis of Akaike's information criterion (AIC). 

In the mixture models, all models included age group and gender as additive covariate effects on the (multinomial) logit component membership probability.  The component-specific time-to-event model was selected by starting with a generalised gamma distribution with constant parameters, then comparing against simpler gamma, Weibull and log-normal models, and against more complex models where one or more of the generalised gamma parameters also depended on age and gender.

In the cause-specific hazards models, we started with generalised gamma distributions for each cause-specific hazard, with a location parameter that depended on age and gender.  This model was then compared with simpler gamma, Weibull and log-normal models, and more complex models where the second or third parameter of the generalised gamma also depended on age and gender.   Models with a ``cure fraction'' (potentially depending on age and gender) were also investigated to describe a proportion
of people at negligible risk of ICU admission or death.  A cure fraction for discharge as well was judged to be implausible, so that while some people will never go to ICU or die from their current infection, all survivors are assumed to eventually leave hospital. 

Interactions between age and gender were also investigated in both frameworks.

\section{Application of the models to the COVID-19 hospital data}
\label{sec:results}

\subsection{Model checking}

The best-fitting among the mixture models and cause-specific hazards models, judging from AIC, are described in Table~\ref{tab:models}.  The best fitting cause-specific hazards model has a lower AIC compared to the best-fitting mixture model, which is driven by the better fit of the cause-specific model for the events following hospital.  After investigating for interactions, effects of age and sex were included as additive in all models.

\begin{table}
\begin{tabular}{lllll}
\hline 
& \multicolumn{2}{l}{\textbf{Cause-specific hazards}} & \multicolumn{2}{l}{\textbf{Mixture multistate}}\\
Transition & Distribution & Covariates on  & Distribution & Covariates on  \\
\hline 
\textbf{From hospital}: & \multicolumn{2}{c}{(AIC 37463)} & \multicolumn{2}{c}{(AIC 38514)}  \\
To ICU & Log-normal cure & $p$ & Log-normal & $\mu$  \\ 
To death & Generalised gamma cure & $p$, $\mu$ &  Generalised gamma & $\mu$  \\ 
To discharge & Generalised gamma & $\mu,\sigma$ & Generalised gamma & $\mu$  \\ 
\hline 
\textbf{From ICU}: & \multicolumn{2}{c}{(AIC 8355)} & \multicolumn{2}{c}{(AIC 8348)}  \\
To death  & Generalised gamma cure & $p$, $\mu$ & Generalised gamma& $\mu$  \\
To discharge & Generalised gamma & $\mu$ & Generalised gamma & $\mu$  \\
\hline 
Total AIC  & \multicolumn{2}{c}{45817 (58 parameters)} & \multicolumn{2}{c}{46862 (52 parameters)}  \\
\hline 
\end{tabular}
\caption{Selected parametric assumptions for the cause-specific hazards and mixture multistate models.  All mixture multistate models also include age and gender covariates on the probabilities $\pi_{rs}$.\label{tab:models}}
\end{table}
The fit of the models can also be compared for specific subgroups of the data.  Table \ref{tab:subgroupfit} shows the difference between the log-likelihood of the cause specific hazards model and the likelihood of the mixture model, for each age/sex subgroup of the data, and for the hospital and ICU-specific submodels and both combined.  The better fit of the cause-specific hazards model overall is due to its better fit in the
the hospital-specific submodel, while the two models fit similarly well for transitions from ICU.   In the hospital submodel, within each subgroup, the log-likelihood of the cause-specific hazards model is greater, showing better fit.  While these subgroup-specific comparisons do not account for the difference in model complexity (as in AIC), there is no evidence that the mixture model should be preferred for describing any particular age/sex subgroup.

\begin{table}
\begin{tabular}{lllll}
\hline
Age group &  Sex &  From hospital &  From ICU &  Combined\\
\hline
44 and under &  Female &  24.7 &  -1.27 &  23.4\\
44 and under &  Male &  46.2 &  -0.14 &  46.1\\
45-64 &  Female &  54.6 &  -0.37 &  54.2\\
45-64 &  Male &  196.1 &  -0.87 &  195.2\\
65-74 &  Female &  32.4 &  -0.44 &  32\\
65-74 &  Male &  89.3 &  -0.21 &  89.1\\
75-84 &  Female &  7.4 &  -0.05 &  7.4\\
75-84 &  Male &  46.8 &  -0.61 &  46.2\\
85+ &  Female &  9.4 &  0.35 &  9.8\\
85+ &  Male &  24.5 &  0.31 &  24.8\\
\hline
\end{tabular}
\caption{Relative fit of the cause-specific hazards model compared to the mixture model, defined as the log likelihood of the cause-specific hazards model minus the log likelihood of the mixture model, by age/sex subgroup, for the hospital submodel, the ICU submodel and both combined. Positive values indicate better fit for the cause-specific hazards model.}
\label{tab:subgroupfit}
\end{table}
  
The fit of the models can be checked against nonparametric estimates in various ways.   Note that  nonparametric estimates are only available up to the maximum observed follow-up in the data, while the parametric models allow extrapolation beyond that time.

\paragraph{Checking mixture and cause-specific hazard models together} 

\begin{figure}
\includegraphics[height=0.47\textheight]{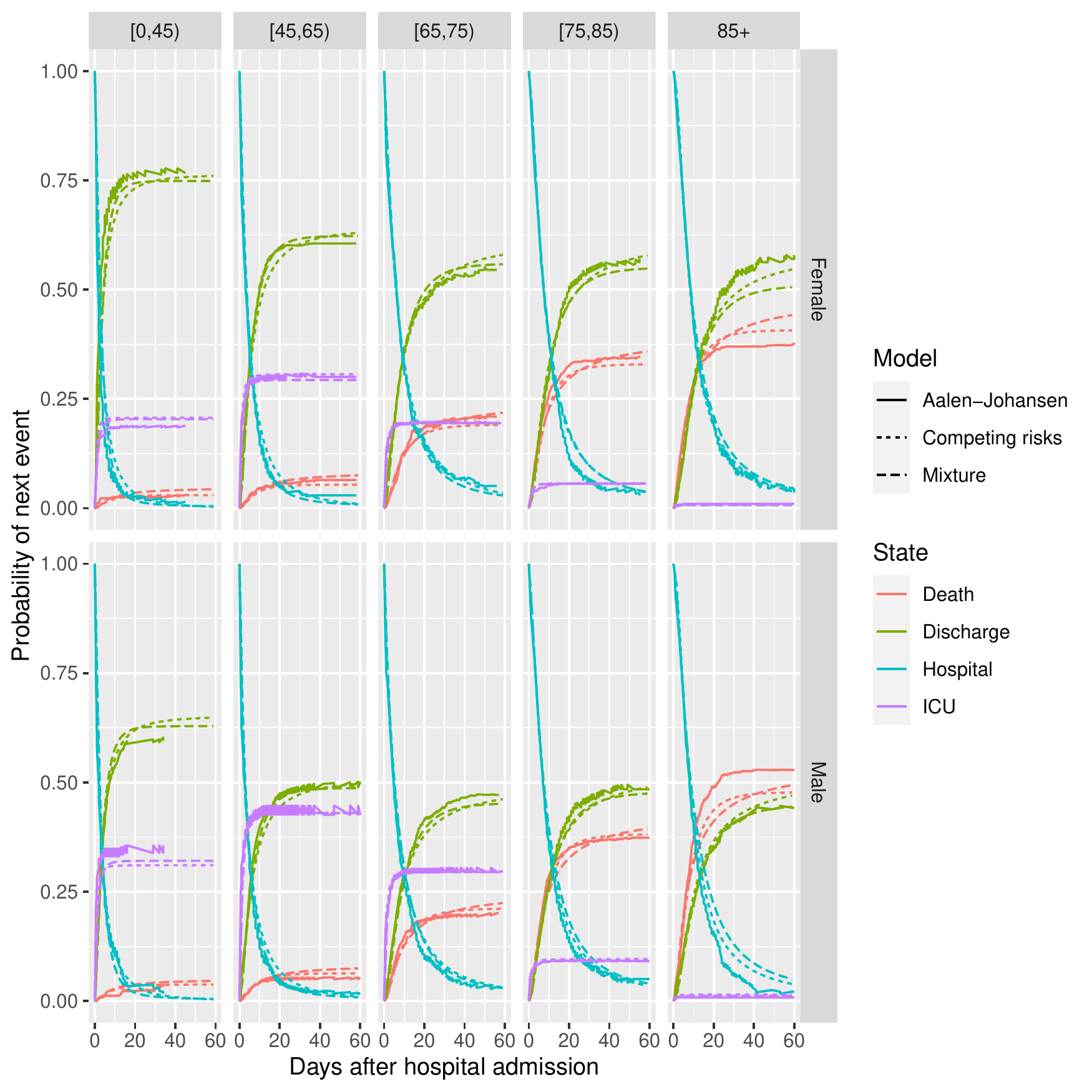}
\includegraphics[height=0.47\textheight]{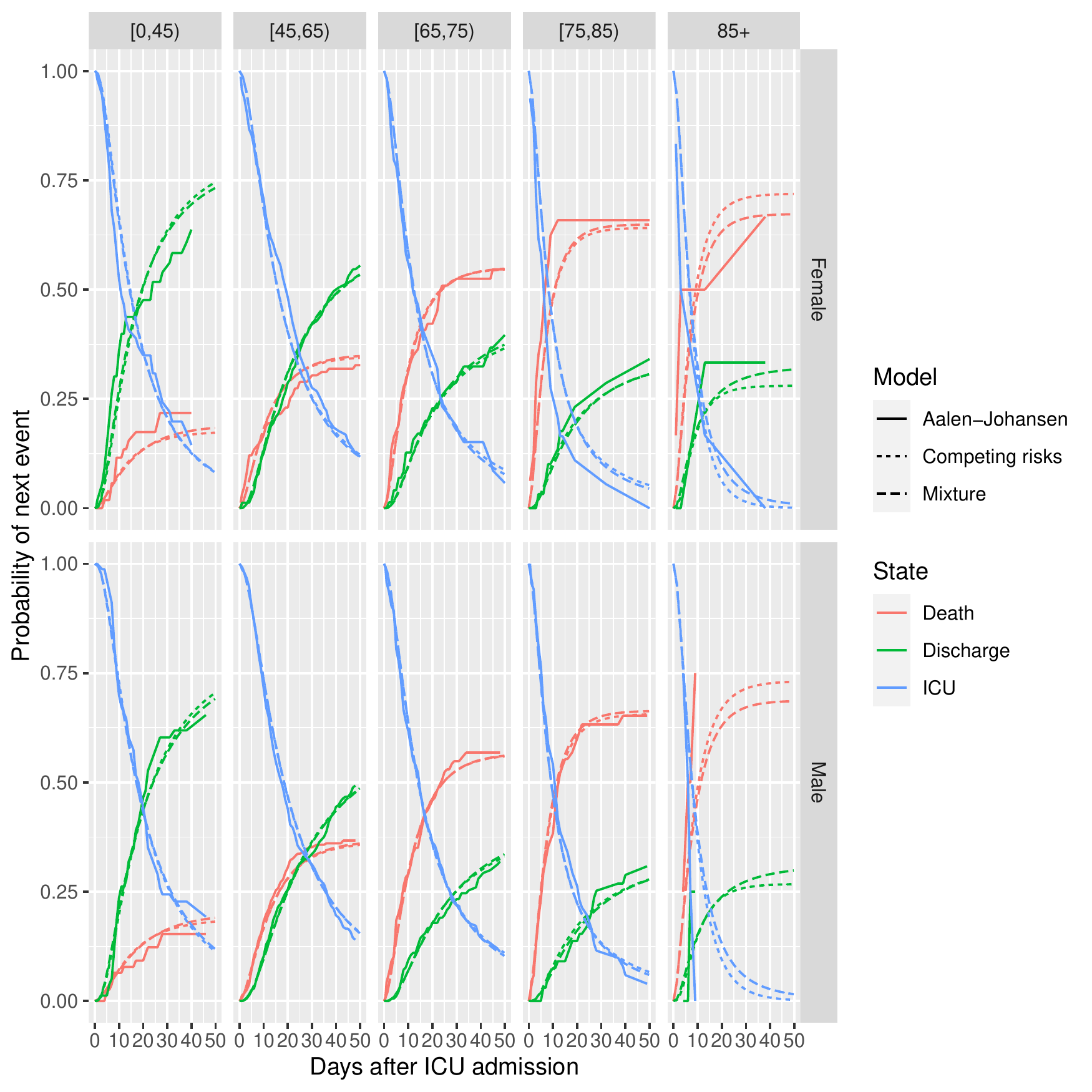}
  \caption[Goodness of fit]{Probabilities of having experienced each competing next state following hospital admission, against time since admission, by age group and gender, comparing non-parametric (Aalen-Johansen) and two alternative parametric models.  Top: events following hospital admission, bottom: events following ICU admission.}
  \label{fig:gof}
\end{figure}

The goodness-of-fit of all of the parametric models can be checked by comparing estimates of the continuous-time transition probability (or ``cumulative incidence'') $p^{(H)}_{r,s}(t),p^{(I)}_{r,s}(t)$  with the standard nonparametric estimates \cite{aalen:johansen}.  Figure~\ref{fig:gof} shows these by age and gender, comparing subgroup-specific Aalen-Johansen estimates against the estimates from the mixture and competing risks models, the upper panel showing the models for $p^{(H)}_{r,s}(t)$, governing the event following hospital admission, and the lower panel showing the models for the event following ICU admission, $p^{(I)}_{r,s}(t)$.    The parametric estimates largely agree with the Aalen-Johansen estimates, except for some disagreement for the mixture model among the oldest and youngest ages, though note the estimates for the events following ICU admission for the older age groups are based on a small sample of ICU admissions for these groups (see Figure~\ref{fig:ridge}).

\paragraph{Check of cause-specific hazard models} 

\begin{figure}
 \includegraphics[width=\textwidth]{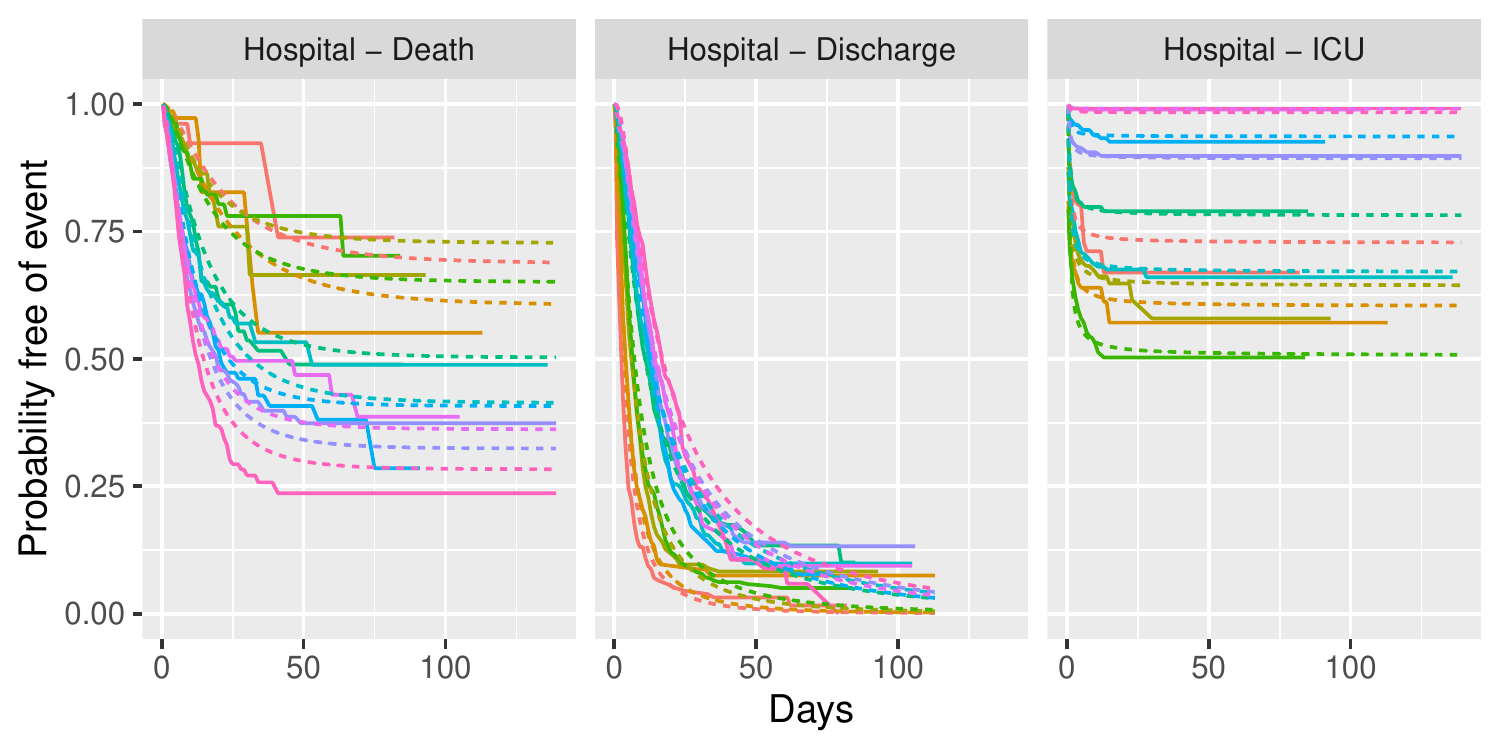}
 \includegraphics[width=\textwidth]{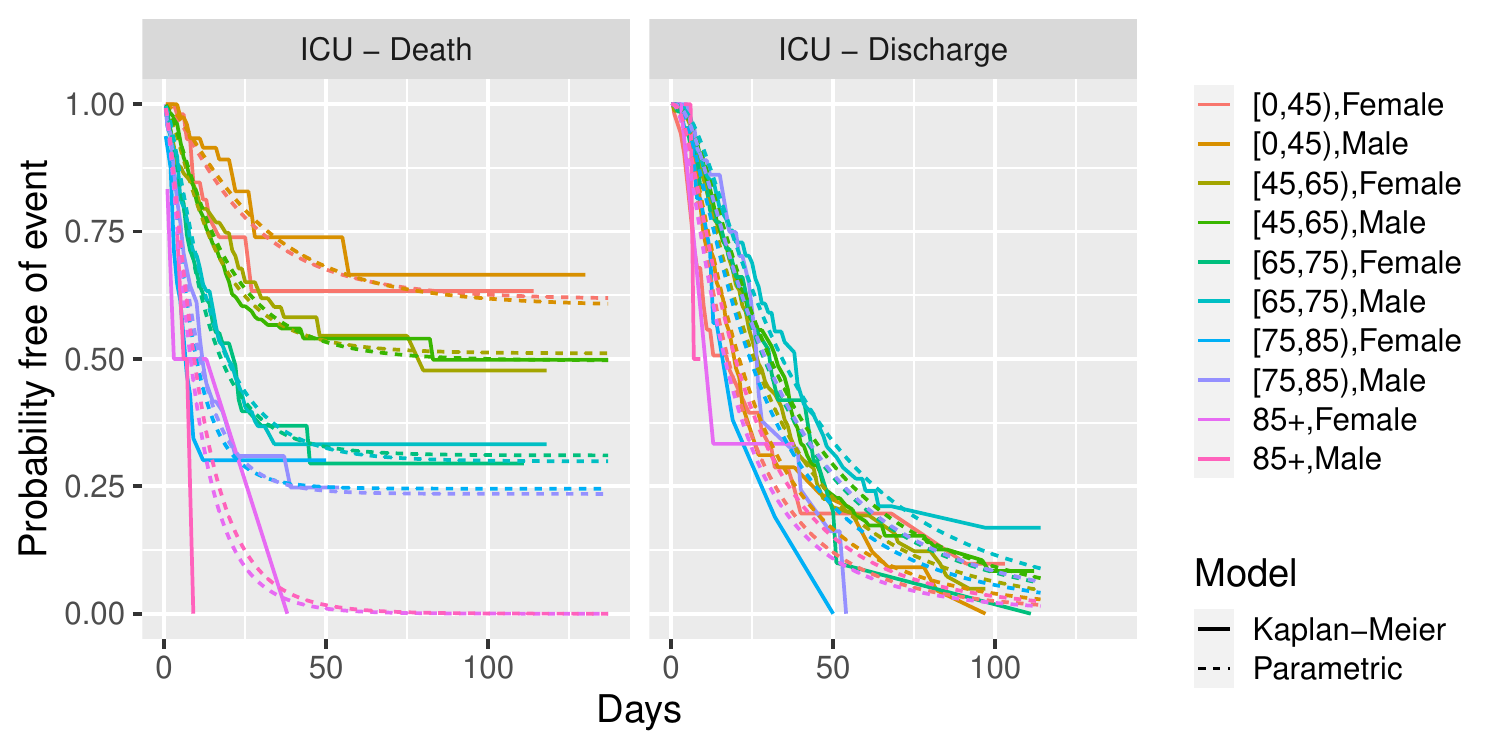}
  \caption[]{Check of cause-specific hazard models against Kaplan-Meier estimates of the distribution of the time to each latent competing event, by age and gender}
  \label{fig:km}
\end{figure}

The cause-specific hazard models can be checked against Kaplan-Meier estimates of the distribution of the time to each latent competing event, since they are standard parametric survival models for the time to the cause of interest, with the occurrence of other causes treated as censoring (Figure~\ref{fig:km}).  

The Kaplan-Meier estimates show a characteristic flattening out for the time from hospital admission to death and ICU admission, implying that only a proportion of patients experience these events by a certain time, after which the hazard of those events becomes small.   This pattern is consistent with the ``cure'' models that provided the best fit among the cause-specific models for these events. 

For the times from hospital to discharge, the Kaplan-Meier estimates show that the majority of (``latent'') discharge times have occurred by around 50 days, if deaths and ICU admissions are considered as right-censoring for these ``latent'' times.  We disregarded ``cure'' models for this transition, assuming that the survival curve for time to discharge will eventually decrease to zero.    However there is a tail of people who have been in hospital for very long periods, particularly from the older age groups (see also Figure~\ref{fig:ridge}).    Therefore we might expect some uncertainty in estimating the upper tail of the distribution of the length of stay.

\paragraph{Check of mixture models} 

\begin{figure}
 \includegraphics[height=0.48\textheight]{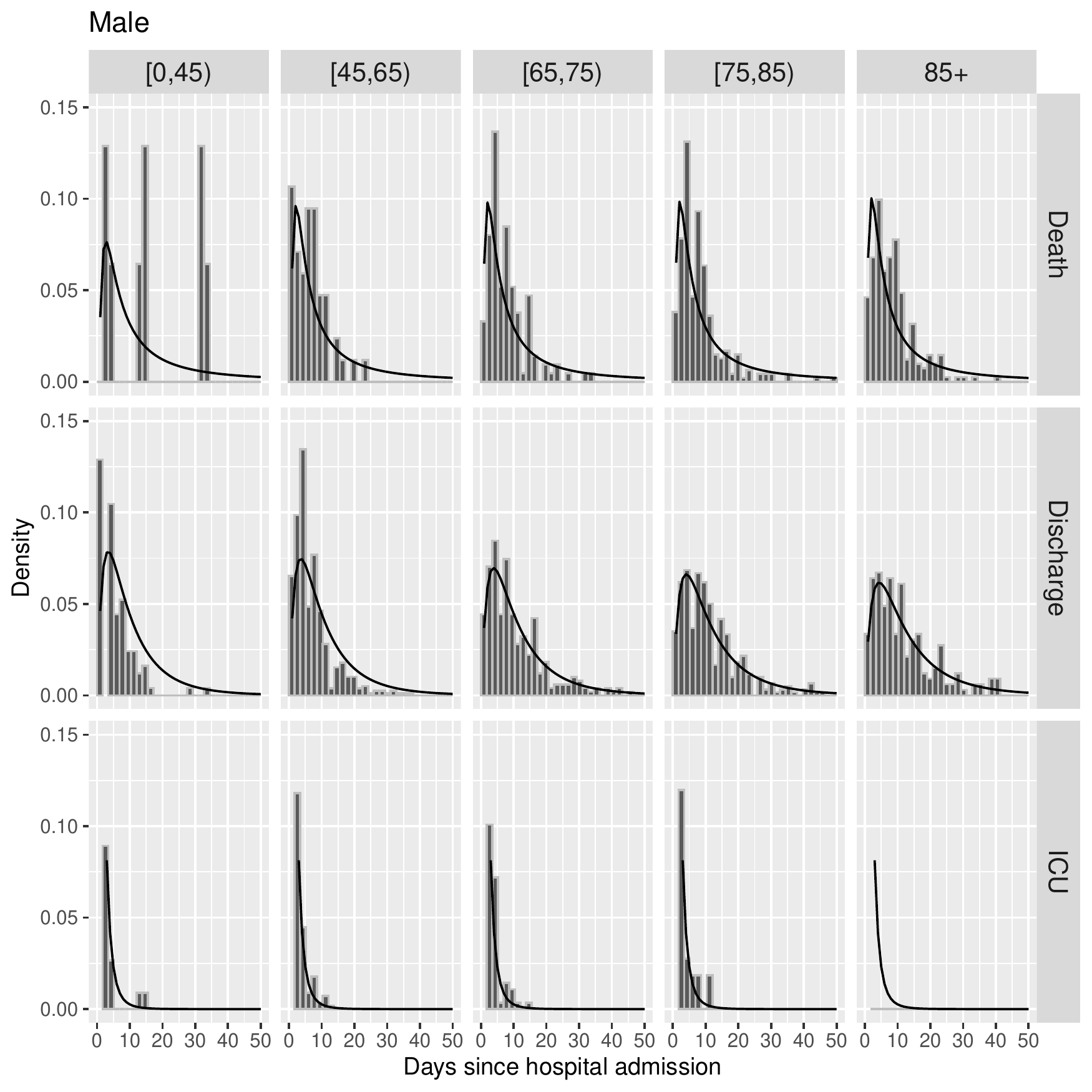}
 \includegraphics[height=0.48\textheight]{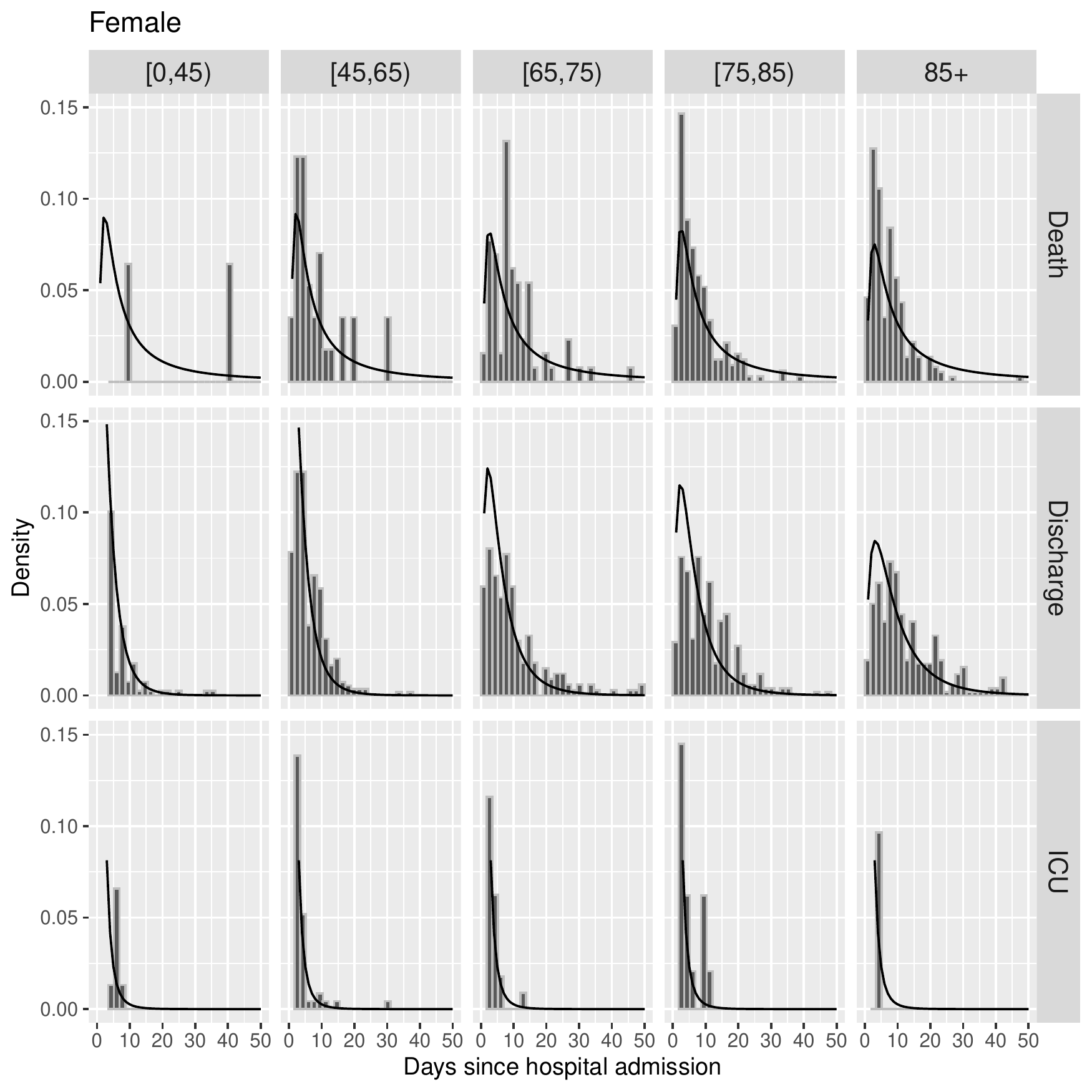}
  \caption[]{Check of fitted time-to-event densities from the mixture model against histograms of observed times to each alternative next event following hospital admission, by gender and age group}
  \label{fig:mixcheckhosp}
\end{figure}

In the mixture models, the fit of the distribution of the time to each event conditionally on that event occurring can be checked, to some extent, against histograms of the observed times.  This ignores the contribution of the censored data to the fitted models, which comprise 10\% of the observations and 
include people with longer hospital stays.  Figure~\ref{fig:mixcheckhosp} shows these comparisons for the model for events following hospital admission, by age and gender.   Estimated densities are overlaid on histograms of the times to each event for people who were observed to have that event.  The shapes of the fitted densities for beyond 50 days of observed follow-up are influenced by the parametric model form, and are difficult to check against the censored and incomplete data (shown in Figure~\ref{fig:ridge}, but not in Figure~\ref{fig:mixcheckhosp}) that contribute to the estimates for these times. 

\subsection{Estimated probabilities and times of next events}

\begin{figure}[htbp]
\includegraphics[]{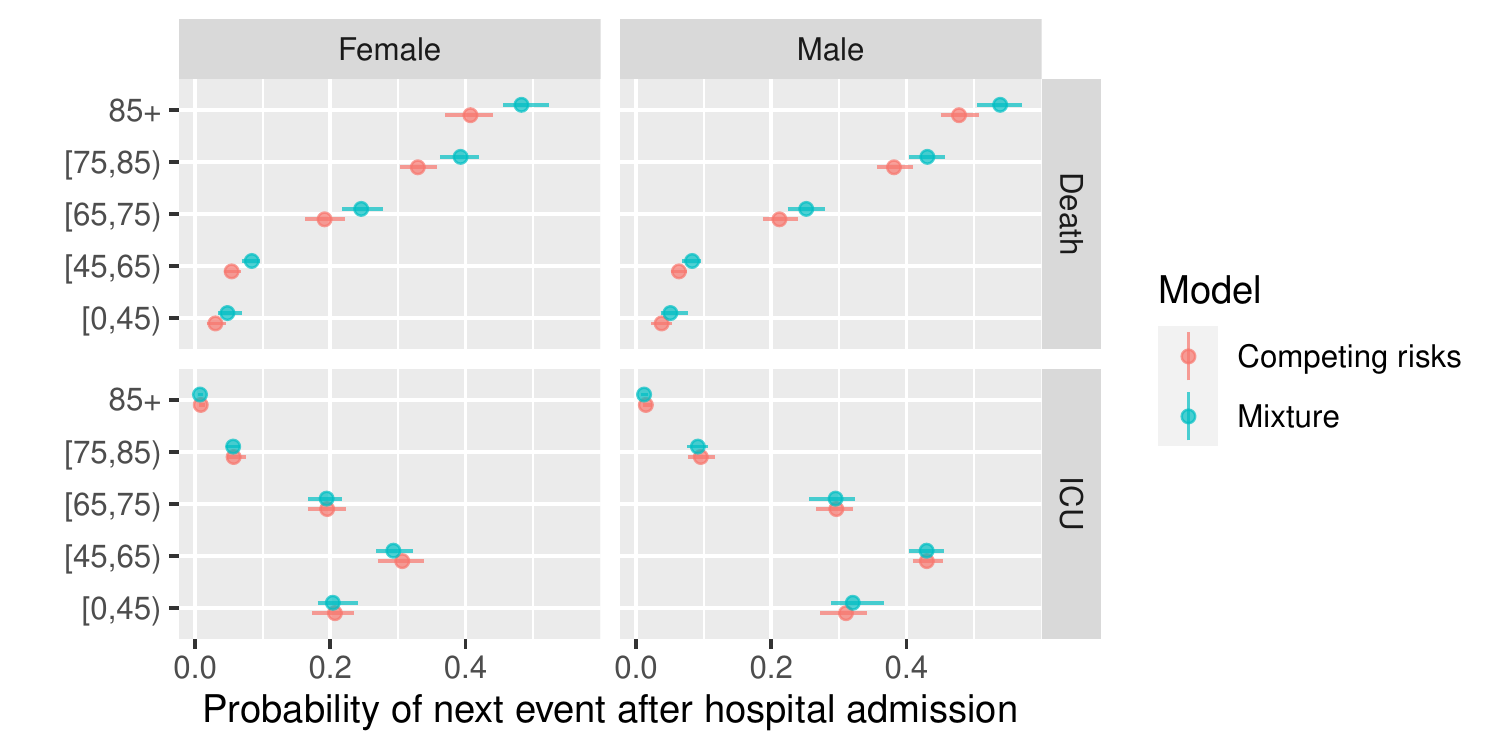}
\includegraphics[]{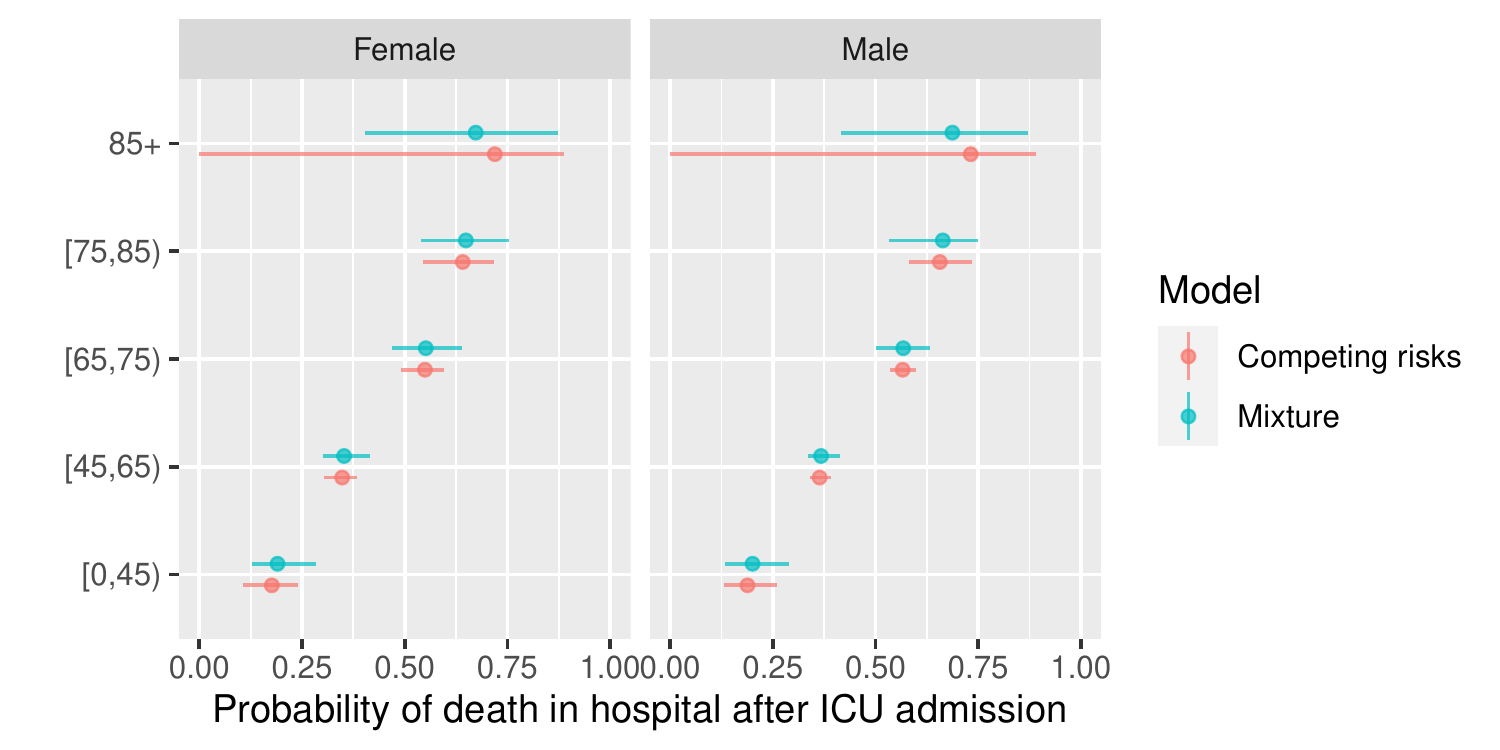}
  \caption{Probabilities of next event after hospital or ICU admission, estimated from mixture and competing risks models.}
    \label{fig:phosp}
\end{figure}

Figure~\ref{fig:phosp} shows the probabilities governing the next event following hospital admission, compared between the two parametric models. There are only moderate differences between the estimates from the mixture and cause-specific hazards models.  The mortality rates, both before and after ICU admission, increase with age and are higher for men.  Rates of ICU admission are higher for younger people, which is largely a consequence of hospital policy rather than disease severity. 

\begin{figure}[htbp]
\includegraphics[]{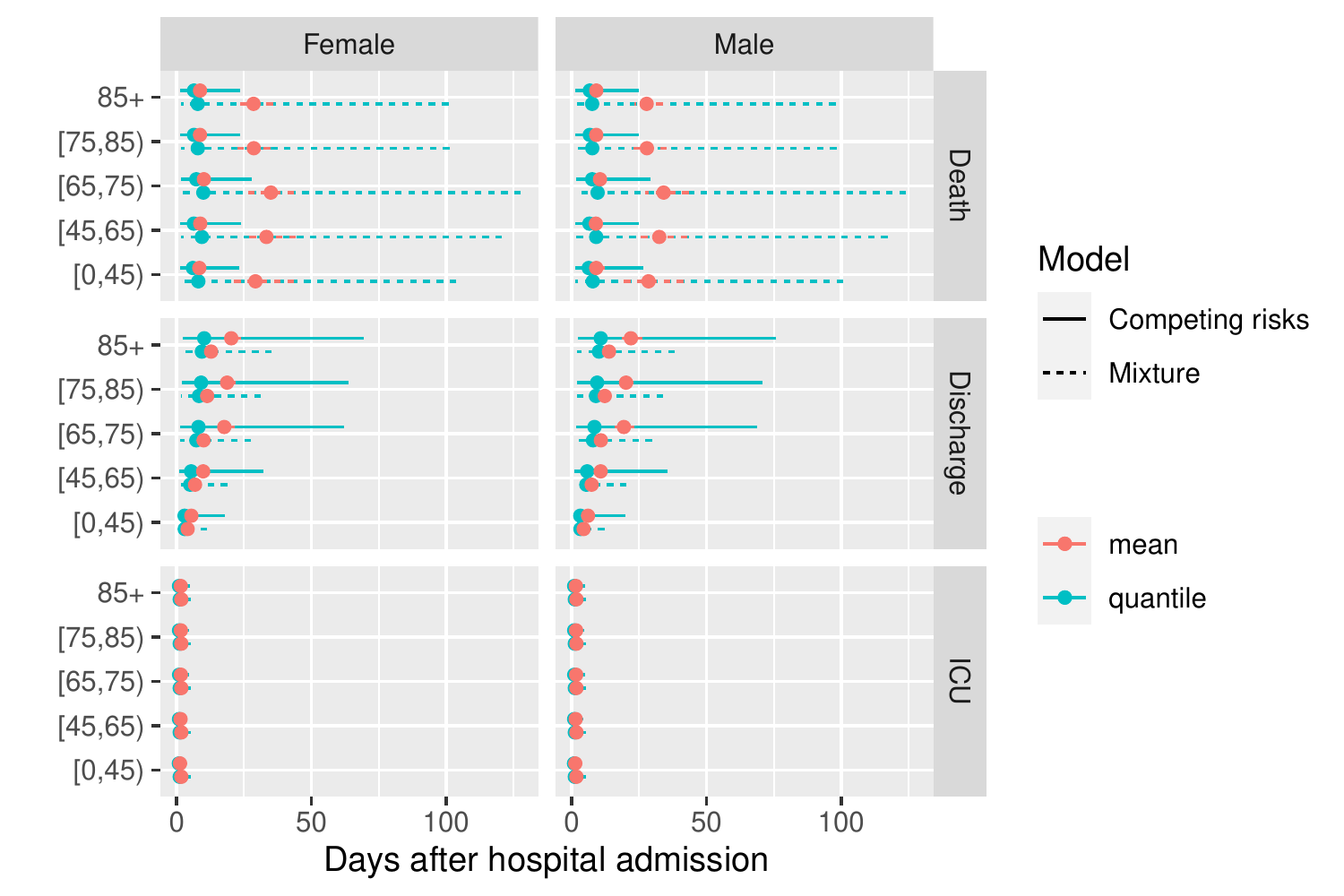}
\includegraphics[]{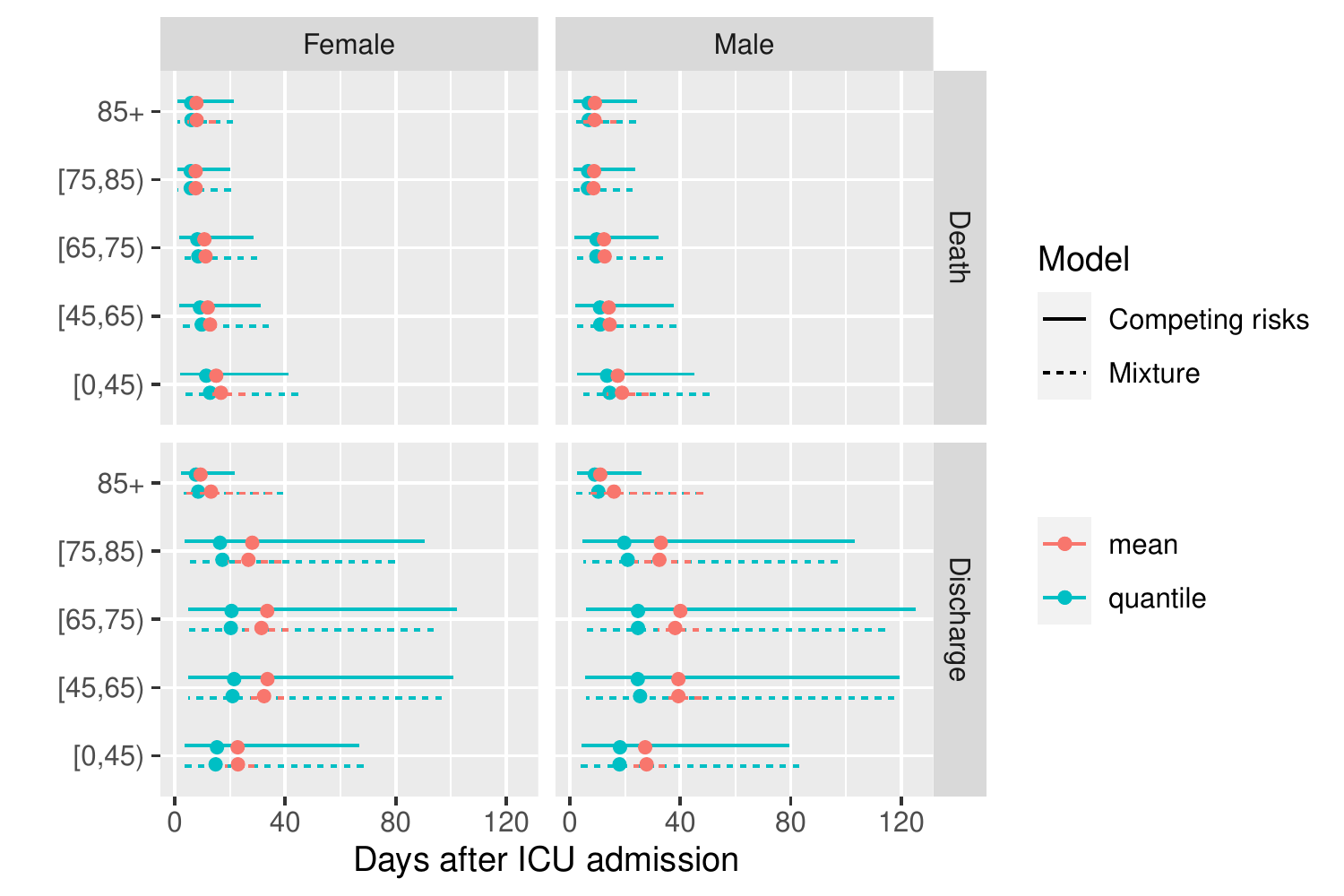}
  \caption{Times to the next event after hospital admission and the next event after ICU admission: means and 95\% confidence intervals in red, and median and 90\% quantile intervals (in blue)}
    \label{fig:meanhosp}
\end{figure}

Figure~\ref{fig:meanhosp} shows the mean times to events (in red) with 95\% confidence intervals (representing uncertainty) and the median and 90\% quantile intervals (in blue, representing between-person variability) for the time-to-event distributions, under both models.    These confidence intervals, and all others presented, are obtained by simulating a sample of alternative parameter values from the asymptotic normal distribution of the maximum likelihood estimates~\cite{mandel2013simulation}.   Note that some of the confidence intervals around the means are too narrow to be seen.  The estimated medians agree between the models, however the estimated upper tails of the distributions are sensitive to the model choice.   The mixture models estimate longer mean times to death (and upper quantiles of this distribution) for those who die in hospital without going to ICU.  This is a plausible consequence of using a cure distribution for time to death in the cause-specific hazard models --- where if a person survives longer than a certain time, the model infers that they belong to the ``cured'' fraction who will never die in hospital.   Whereas under the mixture models, those observed to be still in hospital are estimated to be still at non-zero risk of death, so that longer, but finite, times to death are more plausible under the mixture models considered here.   

Estimated times to discharge are slightly higher under the cause-specific hazards models, which could be explained by artefacts of how the different parametric assumptions extrapolate the time of the eventual event for those who are still in hospital (see also the Kaplan Meier plots of time from hospital to discharge in Figure~\ref{fig:km}).   The models both agree on the time from hospital to ICU admission being distributed tightly around a mean of about 2 days, and on the distributions of the times to events following ICU admission. 

\subsection{Estimates for ultimate events after hospital admission}

\begin{figure}[htbp]
\includegraphics[width=\textwidth]{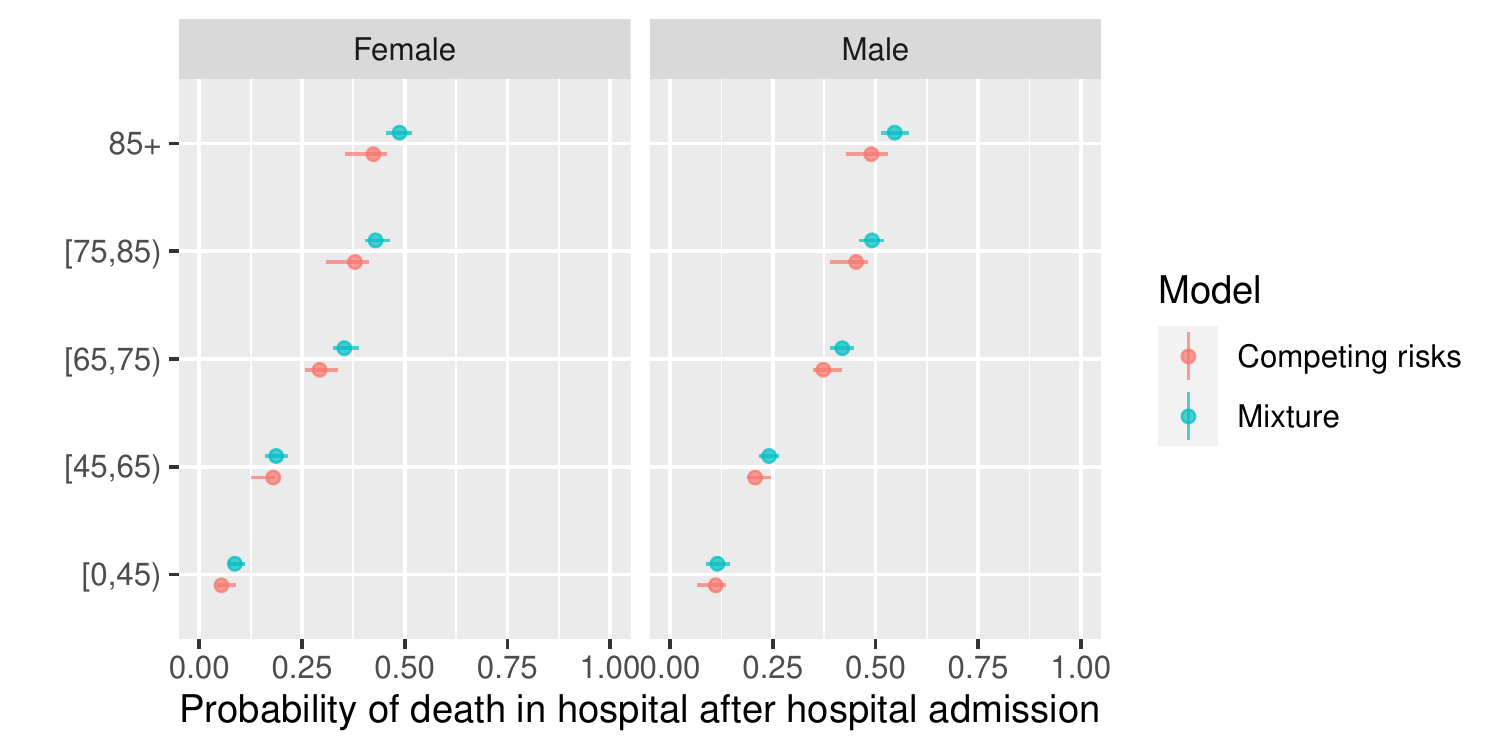}
\includegraphics[width=\textwidth]{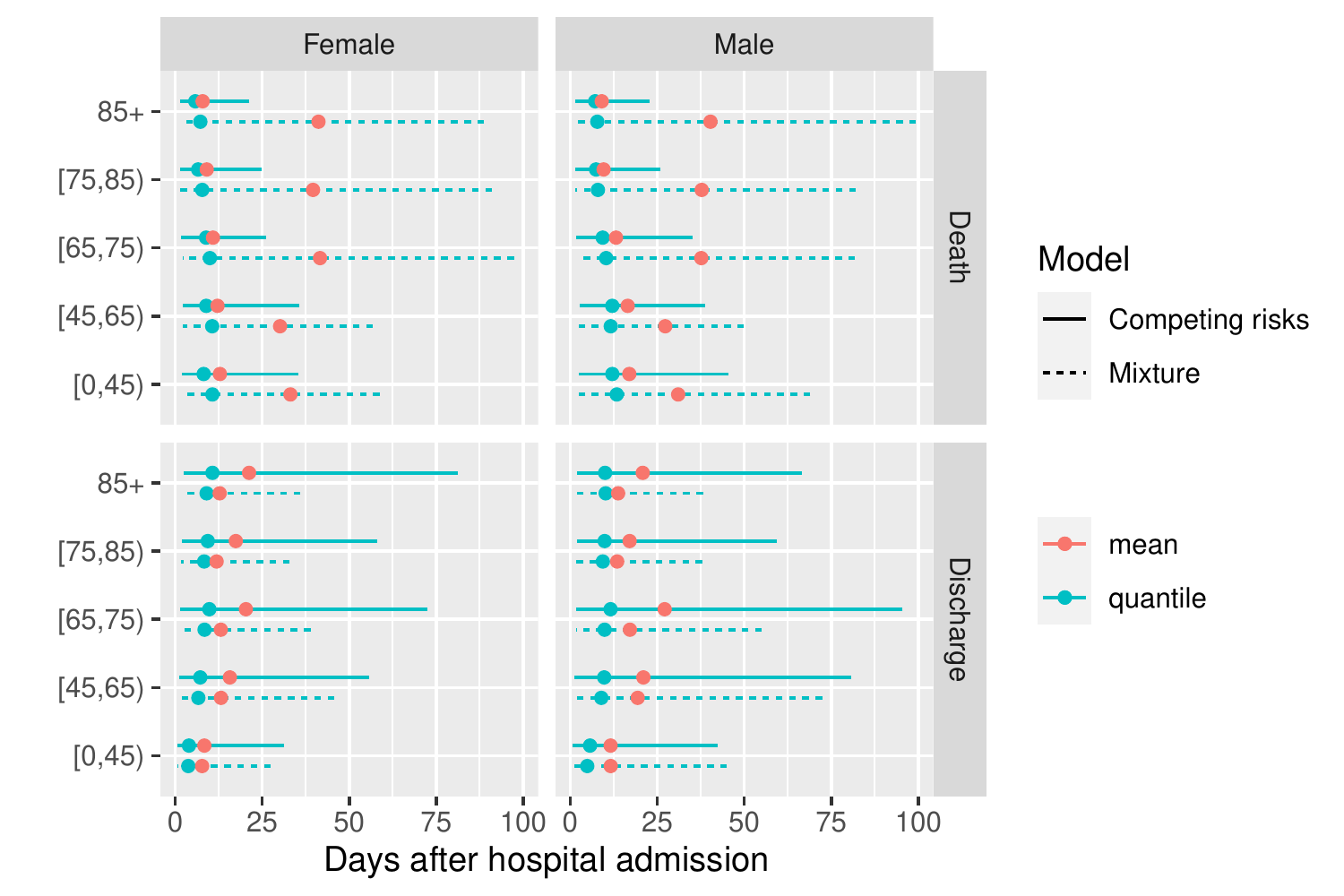}
  \caption{Probabilities of death in hospital, and times from hospital admission to death in hospital or discharge, averaged over people who are admitted to ICU and people who aren't.}
    \label{fig:final}
\end{figure}

The models for the events following hospital admission and events following ICU admission can be coupled to provide predictions of the probability that a person just admitted to hospital will die in hospital, and the distributions of the time to death in hospital or discharge alive (Figure~\ref{fig:final}).     The estimated probabilities are again slightly different between the two models, but both show the same increasing trend with age, and higher mortality for men, from around 0.07 for men and women aged under 45 to 0.4 for women over 85 years of age, and 0.5 for men over 85. 

The differences between models in the estimated distribution of times to ultimate events reflect the differences seen in  Figure~\ref{fig:meanhosp}, with medians that agree between the models (at around 10 days for times to death and discharge), but longer mean times to death under the mixture model, and longer mean times to discharge under the competing risks model. 

While the cause-specific hazards model fitted the observed data better overall (judging by AIC), inference from the fitted mixture model is more computationally efficient.  Computing the quantities of interest is practically instant given the fitted mixture model.    Computing them given the competing risks model requires two levels of simulation: over $S$ individuals to determine the sampling distribution of the minimum, and over $B$ alternative parameter values to represent parameter uncertainty.   Using $S=100000$ and $B=100$ to simulate the required quantities in this application took around 8 minutes on a single computer processor, though the computation would be simple to parallelise.

\section{Discussion and comparison of modelling frameworks}
\label{sec:discussion}

We have obtained estimates of probabilities of events following hospital admission, and estimates of the
distribution of times to those events, for people admitted to hospital with COVID-19 infection, using two
different frameworks for multi-state modelling.    The estimates of event probabilities and median times to events did not depend substantially on the model assumptions.  However, due to limited follow up, and despite the fact that only about 10\% of event times were censored, some uncertainty remained in the upper tails of the distributions of the times to death and discharge.    The parametric assumptions can only be checked against the data observed in the follow-up period, where a cause-specific hazard model with ``cure'' fractions was found to fit best, based on AIC and comparisons with nonparametric estimates.    To make longer-term predictions we must make substantive assumptions about what will happen after the end of follow up.   In this example, we assumed ``cure'' fractions for ICU and death were plausible in the cause-specific hazards model, that is, a proportion of people never experience these events, but everyone will eventually be discharged from hospital. 

Both  mixture modelling  and cause-specific hazards are useful frameworks for fully-parametric multi-state modelling.  With the \texttt{flexsurv} R package, they can both be applied to general multi-state data with a wide range of flexible parametric distributions and covariate dependencies.    While \emph{fitting} the mixture model is more demanding due to the need to maximise over a larger number of parameters, easily-interpretable quantities, such as probabilities of and and times to observed events, are easier to compute under the mixture model.  Fit of both models can be checked against nonparametric estimates. 

As discussed by Cox~\cite{cox1959analysis}, in theory, either model framework can represent the exact mechanism if the transition-specific parametric distributions are specified correctly.  A third framework, ``vertical''  modelling, was also proposed by Nicolaie et al. \cite{nicolaie2010vertical}, based on modelling
$P(time)P(event|time)$, rather than $P(event) P(time|event)$ as in the mixture model.  Though in practice, very flexible model families and large samples would often be required for the best-fitting model among different frameworks to give substantively identical inferences.  For example, suppose there were two competing latent event times $T_1$, $T_2$ with cause-specific hazards distributed as Gamma$(a_1,b_1)$ and Weibull$(a_2,b_2)$ respectively, then the equivalent mixture model would be specified by the conditional distributions of $T_1|T_1<T_2$ and $T_2|T_1 > T_2$, which wouldn't have a standard form.   In principle, splines \cite{royston:parmar,crowther2014general,komarek2005accelerated} might be used to construct arbitrarily-flexible time-to-event models --- there is general-purpose software for these, though identifiability and computational challenges may constrain the flexibility they allow in practice.   

Either model framework might be used in situations where some people are at negligible risk of particular events, as in ``cure'' models where a fraction of people do not die from a disease.   We clarified the difference between the ``mixture cure'' distributions of Boag \cite{boag1949maximum} and the ``mixture competing risks'' models of Larson and Dinse \cite{larson1985mixture} --- in the former, the ``cure'' event that competes with death is not observable, and in the latter, it is.  We extended the mixture competing risks  model to a full multi-state model, and  developed accessible software to implement it, while we used the mixture cure distribution to define cause-specific hazards in the competing risks framework.   This provided the best fit to the COVID-19 hospital data, judging by AIC.   This might have been because the ``mixture competing risks'' model does not assume that a person in one mixture component is ``immune'' from the events defining the other components --- because the components are simply defined by which event among a set of competing events will occur before the others.    In our application, the cause-specific cure model, where a proportion of people are at zero risk of ICU admission and death in hospital, fitted better than the mixture model where everyone still in hospital is assumed to be still at risk of these events.

In our application, policy-makers required estimates of average outcomes for mixed populations defined by age groups and gender.  Therefore there were only two categorical covariates in our model.  Models with more covariates, including continuous covariates, would be required to determine predictors of outcomes for individuals, or to investigate disease aetiology.   Under the mixture model, covariate effects on probabilities of, or times to, observable events can be estimated directly, e.g. as log odds ratios, hazard ratios or time acceleration factors.  While hazard ratios from a cause-specific hazard model can be argued to more closely represent the mechanisms of how risks of events are determined at the biological level, compared to effects on probabilities \citep{pintilie2006competing}, they are harder to interpret in terms of average outcomes compared between populations.   Non-proportional hazards, or other flexible models for covariate dependencies, are available in software (e.g. \texttt{flexsurv}), and any quantity of interest can be computed and contrasted between specific covariate values, however this may require expensive simulation.   Another approach to obtaining ``average effects'' of specific covariates involves regression on the cumulative incidence function, which is possible in a fully parametric framework~\citep{lambert2017flexible}, as well as semi-parametrically~\citep{fine1999proportional}.    With continuous covariates or many covariates, goodness of fit checking would also be more challenging, compared to our model where we simply compared model predictions with stratified nonparametric estimates.      Also with many covariates, the computational advantages of the cause-specific hazards model, in terms of fitting, would become more apparent --- since the cause-specific likelihood factorises into independent components, compared to the mixture model which involves a joint likelihood over all competing events. 

In routinely-collected data on people hospitalised with an infection, other challenges might arise, such as more severe kinds of incomplete observation.   Addressing these challenges would be important for decision-making at the start of an epidemic, where data are sparser.    In earlier versions of the dataset that we studied, final survival status and ICU admission histories were missing for substantial numbers of patients, and many event dates were interval-censored over wide ranges.   Such partial observations are hard to handle without strong assumptions such as the Markov assumption and piecewise-constant hazards~\citep{kalbfleisch:lawless}, and even with a flexible model, untestable assumptions about whether missingness is informative may be required.   Routinely-collected data are also subject to selection biases which may make inference for wider populations difficult.   These challenges further emphasise the need for strong infrastructures for data collection in preparation for future public health emergencies. 

Code to implement the analyses in the paper is included as an online supplementary document in R Markdown format, together with a simulated dataset of the same form as the data used in the paper.

\paragraph{Acknowledgements}
This work was supported by the Medical Research Council programmes MRC\_MC\_UU\_00002/11, MRC\_MC\_UU\_00002/10 and MRC\_MC\_UU\_00002/2.   We are grateful to the Joint Modelling Cell and Epidemiology Cell at Public Health England for providing and discussing the CHESS data.

\bibliography{covid_multistate}

\end{document}